\documentclass[twocolumn,showpacs,preprintnumbers,showkeys,superscriptaddress]{revtex4}
\usepackage{amssymb}
\usepackage{graphicx}
\usepackage{dcolumn}
\usepackage{bm}
\usepackage{appendix}

\def\eqref#1{Eq.~(\ref{eq:#1})}

\begin{document}

\title{Particle-Hole Symmetry in Generalized Seniority, Microscopic Interacting Boson (Fermion) Model, Nucleon-Pair Approximation, and Others}
\author{L. Y. Jia}  \email{liyuan.jia@usst.edu.cn}
\affiliation{Department of Physics, University of Shanghai for
Science and Technology, Shanghai 200093, P. R. China}

\date{\today}

\begin{abstract}

The particle-hole symmetry (equivalence) of the full shell-model Hilbert space is straightforward and routinely used in practical calculations. In this work we show that this symmetry is preserved in the subspace truncated at a certain generalized seniority, and give the explicit transformation between the states in the two types (particle and hole) of representations. Based on the results, we study the particle-hole symmetry in popular theories that could be regarded as further truncations on top of the generalized seniority, including the microscopic interacting boson (fermion) model, the nucleon-pair approximation, and others.

\end{abstract}


\vspace{0.4in}

\maketitle

\section{Introduction}

The particle-hole symmetry (equivalence) of the full shell-model Hilbert space is straightforward. A Slater determinant of $2 N$ particles is the same state as a Slater determinant of $2(\Omega - N)$ holes within a model space of degeneracy $2\Omega = \sum_j (2j+1)$. Operators should be converted accordingly as in textbooks \cite{Bohr_book, Suhonen_book}, and the final results are independent of whether choosing particles or holes as the degree of freedom. Practical shell-model calculations frequently encounter the dimension limitation, and various truncation schemes are necessary in reducing the dimension. Apparently it is desirable to preserve the particle-hole symmetry when truncating; in this work we consider whether this is the case for some popular truncation schemes.

The seniority quantum number $\nu$ was first introduced by Racah \cite{Racah_1943, Racah_1952, Flowers_1952} as the number of unpaired nucleons in a single $j$-level to incorporate pairing correlations. As a truncation scheme for the realistic multi-$j$ shell model, the seniority $\nu$ equals to the total number of unpaired nucleons in all $j$-levels. Obviously, a $2N$-particle Slater determinant of seniority $v$ is the same state as a $2(\Omega-N)$-hole Slater determinant of seniority $v$. The particle-hole symmetry is preserved in the seniority truncated subspaces.

The generalized seniority quantum number $S$ was also introduced \cite{Talmi_1971, Shlomo_1972, Gambhir_1969, Gambhir_1971, Allaart_1988, Talmi_book} as the number of unpaired nucleons in a multi-$j$ model; but the paired part wavefunction is uniquely written as the condensate of coherent pairs. The generalized seniority states are no longer Slater determinants, and the particle-hole symmetry is not obvious. Using commutator techniques in J-scheme, Talmi showed \cite{Talmi_82} that the $2N$-particle state of $S=0$ is the same as the $2(\Omega-N)$-hole state of $S=0$ with reciprocal coherent pair structures; and the $S=2$ particle states span the same subspace as the $S=2$ hole states. But for $S > 2$ the conclusion is absent. Along the same line the particle-hole symmetry found by Johnson and Vincent \cite{Johnson_85} is restricted within the ${\mathcal{S}}$-${\mathcal{D}}$ subspace; and their collective quadrupole pair operator ${\mathcal{D}}$ is defined with generalized-seniority projection thus is different from the usual one. Ref. \cite{Johnson_85} was mainly written for the microscopic foundation of the interacting boson model. The particle-hole symmetry for arbitrary generalized seniority $S$ was claimed in Ref. \cite{Allaart_1988} but without a proof; in fact, their only results for $S=2$ [Eqs. (2.44) and (2.95)] were misprinted.

In this work we show that the particle-hole symmetry exists for arbitrary generalized seniority $S$. In addition, we give the explicit transformation of states between the particle and the hole representations, in both the M-scheme and the J-scheme. Based on the results, we consider the particle-hole symmetry for popular theories that could be regarded as further truncations on top of the generalized seniority, including the microscopic interacting boson (fermion) model, the nucleon-pair approximation, and others.

Section \ref{sec_M_scheme} discusses in M-scheme the particle-hole symmetry in generalized seniority. The M-scheme results are coupled into J-scheme expressions in Sec. \ref{sec_J_scheme}. We consider in Secs. \ref{sec_IBM}, \ref{sec_NPA}, and \ref{sec_others} the particle-hole symmetry in the microscopic interacting boson (fermion) model, the nucleon-pair approximation, and other popular truncation schemes. Section \ref{sec_conclusions} summarizes the work.

\section{M-Scheme Generalized Seniority  \label{sec_M_scheme}}

In this section we show that the particle-hole symmetry exists in generalized-seniority truncated subspaces. The time-reversal invariance is assumed but not necessarily the rotational symmetry; hence the results are valid for deformed (Nilsson) single-particle levels. Briefly reviewing definitions of generalized seniority, the
pair-creation operator
\begin{eqnarray}
P_\alpha^\dagger = a_\alpha^\dagger a_{\tilde{\alpha}}^\dagger  \label{P1_dag}
\end{eqnarray}
creates a pair of particles on the single-particle level $|\alpha\rangle$
and its time-reversed partner $|\tilde{\alpha}\rangle$
($|\tilde{\tilde{\alpha}}\rangle = - |\alpha\rangle$, $P_\alpha^\dagger = P_{\tilde{\alpha}}^\dagger$). The coherent
pair-creation operator
\begin{eqnarray}
P^\dagger = \sum_{\alpha \in \Lambda} v_\alpha P_\alpha^\dagger  \label{P_dag}
\end{eqnarray}
creates a pair of particles coherently distributed with structure
coefficients $v_\alpha$ over the entire single-particle space, where the summation index $\alpha \in \Lambda$ runs over the ``pair index'' space $\Lambda$ that is half of the single-particle space (for example,
only those orbits with a positive magnetic quantum
number $m$). The unnormalized pair-condensate wavefunction of the $2N$-particle system
\begin{eqnarray}
(P^\dagger)^{N} |0\rangle
 \label{gs}
\end{eqnarray}
builds in pairing correlations. Gradually breaking coherent pairs in the pair condensate (\ref{gs}), the generalized seniority ($S=2s$) is introduced \cite{Talmi_1971, Shlomo_1972, Gambhir_1969, Gambhir_1971, Allaart_1988, Talmi_book} as the number of unpaired particles
\begin{eqnarray}
\underbrace{a^\dagger a^\dagger ... a^\dagger}_{S =
2s} (P^\dagger)^{N-s} | 0 \rangle .
\label{sen_basis}
\end{eqnarray}
Practical calculations usually truncate the full many-body space at a certain $S$ ($S = 2N$ corresponds to the full space without truncation).

The full space has the particle-hole symmetry. Assuming the single-particle space has degeneracy $2\Omega$, the Hilbert space that consists of Slater determinants of $2N$ particles is the same as that of $2\bar{N} \equiv 2(\Omega-N)$ holes. Now we consider whether the above two Hilbert spaces truncated at certain generalized seniority $S=2s$ are still the same [$0 \le s \le \min(N,\bar{N})$]. To simplify notations we define
\begin{eqnarray}
\eta^s_{s'} = \frac{(N-s)!}{(\bar{N}-s')!} \sum_{\alpha \in \Lambda} v_\alpha .  \nonumber
\end{eqnarray}

Talmi has shown \cite{Talmi_82} that for $s=0$ they are the same. The $2N$-particle pair condensate (\ref{gs}) is the same state as the $2\bar{N}$-hole pair condensate
\begin{eqnarray}
(\bar{P})^{\bar{N}} |\bar{0}\rangle  \label{gs_ph}
\end{eqnarray}
with reciprocal pair structures
\begin{eqnarray}
\bar{P} = \sum_{\alpha \in \Lambda} \frac{1}{v_\alpha} P_\alpha .  \label{P_bar}
\end{eqnarray}
In Eq. (\ref{gs_ph}),
\begin{eqnarray}
|\bar{0}\rangle = \prod_{\alpha \in \Lambda} P_\alpha^\dagger |0\rangle
\end{eqnarray}
is the completely occupied state (closed shell). This result was re-derived with the correct normalization in Eq. (2.90) of Ref. \cite{Allaart_1988},
\begin{eqnarray}
(P^\dagger)^{N} |0\rangle = \eta ^0_0 (\bar{P})^{\bar{N}} |\bar{0}\rangle .  \label{Talmi}
\end{eqnarray}

For $s=1$, Talmi proved \cite{Talmi_82} the particle-hole symmetry through commutator techniques in (coupled) $J$-scheme. Here we prove it in $M$-scheme by using the identity (\ref{Talmi}) in Pauli-blocked spaces; this proof seems more clear and can be directly generalized to $s \ge 2$. We divide the $s=1$ states into two types: $a_\alpha^\dagger a_\beta^\dagger (P^\dagger)^{N-1} |0\rangle$ (where $\alpha$ and $\beta$ belong to different time-reversal pairs, $P_\alpha \ne P_\beta$) and $a_\alpha^\dagger a_{\tilde{\alpha}}^\dagger (P^\dagger)^{N-1} |0\rangle$. The first type is
\begin{eqnarray}
typeI = a_\alpha^\dagger a_\beta^\dagger (P^\dagger)^{N-1} |0\rangle = (P^\dagger_{[\alpha\beta]})^{N-1} a_\alpha^\dagger a_\beta^\dagger |0\rangle ,  \label{A1}
\end{eqnarray}
where $P^\dagger_{[\alpha\beta]} \equiv P^\dagger - v_\alpha P_\alpha^\dagger - v_\beta P_\beta^\dagger$ is the coherent pair-creation operator removing $P_\alpha^\dagger$ and $P_\beta^\dagger$ due to Pauli blocking. For convenience we introduce $|0^{[\alpha\beta]}\rangle$ to represent a subspace
of the original single-particle space, by removing pairs of single-particle levels $|\alpha\rangle, |\tilde{\alpha}\rangle$ and $|\beta\rangle, |\tilde{\beta}\rangle$ from the latter. Within the subspace $|0^{[\alpha\beta]}\rangle$ the identity (\ref{Talmi}) still holds [the power of $\bar{P}$ corresponding to $(P^\dagger)^{N-1}$ should be $(\Omega-2)-(N-1) = \bar{N}-1$],
\begin{eqnarray}
(P^\dagger_{[\alpha\beta]})^{N-1} |0^{[\alpha\beta]}\rangle  \nonumber \\
= \frac{(N-1)!}{(\bar{N}-1)!} (\prod_{\gamma \in \Lambda}^{\gamma \ne \alpha,\beta} v_\gamma) (\bar{P}_{[\alpha\beta]})^{\bar{N}-1} |\bar{0}^{[\alpha\beta]}\rangle  \nonumber \\
= \frac{\eta^1_1}{v_\alpha v_\beta}  (\bar{P}_{[\alpha\beta]})^{\bar{N}-1} |\bar{0}^{[\alpha\beta]}\rangle ,  \nonumber
\end{eqnarray}
where $\bar{P}_{[\alpha\beta]} \equiv \bar{P} - P_\alpha/v_\alpha - P_\beta/v_\beta$ and $|\bar{0}^{[\alpha\beta]}\rangle = P_\alpha P_\beta |\bar{0}\rangle$. Thus Eq. (\ref{A1}) becomes
\begin{eqnarray}
typeI = \eta^1_1 \frac{a_{\tilde{\alpha}} a_{\tilde{\beta}}}{v_\alpha v_\beta} (\bar{P})^{\bar{N}-1} |\bar{0}\rangle .  \label{A2}
\end{eqnarray}
The result is a $s=1$ state in the hole representation.

The second type is treated similarly,
\begin{eqnarray}
typeII = a_\alpha^\dagger a_{\tilde{\alpha}}^\dagger (P^\dagger)^{N-1} |0\rangle = (P_{[\alpha]}^\dagger)^{N-1} P_\alpha^\dagger |0\rangle .  \label{B1}
\end{eqnarray}
The identity (\ref{Talmi}) in the subspace $|0^{[\alpha]}\rangle$ gives [$(\Omega-1)-(N-1) = \bar{N}$]
\begin{eqnarray}
(P^\dagger_{[\alpha]})^{N-1} |0^{[\alpha]}\rangle = \frac{(N-1)!}{\bar{N}!} (\prod_{\gamma \in \Lambda}^{\gamma \ne \alpha} v_\gamma) (\bar{P}_{[\alpha]})^{\bar{N}} |\bar{0}^{[\alpha]}\rangle  \nonumber \\ 
= \frac{\eta^1_0}{v_\alpha}  (\bar{P}_{[\alpha]})^{\bar{N}} |\bar{0}^{[\alpha]}\rangle .  \nonumber
\end{eqnarray}
Therefore Eq. (\ref{B1}) becomes
\begin{eqnarray}
typeII = \frac{\eta^1_0}{v_\alpha}  (\bar{P}_{[\alpha]})^{\bar{N}} |\bar{0}\rangle .  \label{B2}
\end{eqnarray}
The binomial expansion of $(\bar{P}_{[\alpha]})^{\bar{N}} = (\bar{P} - P_\alpha/v_\alpha)^{\bar{N}} = (\bar{P})^{\bar{N}} - \bar{N} (\bar{P})^{\bar{N}-1} P_\alpha / v_\alpha + ...$ has $\bar{N}+1$ terms; but terms with $(P_\alpha)^2$ or higher powers vanish when acting on $|\bar{0}\rangle$ due to Pauli's principle. Thus
\begin{eqnarray}
typeII = \frac{\eta^1_0}{v_\alpha} (\bar{P})^{\bar{N}} |\bar{0}\rangle + \frac{\eta^1_1}{(v_\alpha)^2} a_{\tilde{\alpha}} a_{\tilde{\tilde{\alpha}}} (\bar{P})^{\bar{N}-1} |\bar{0}\rangle ,  \label{B3}
\end{eqnarray}
where we have used $\eta^1_0 \bar{N} = \eta^1_1$ and $P_\alpha = a_{\tilde{\alpha}} a_\alpha = - a_{\tilde{\alpha}} a_{\tilde{\tilde{\alpha}}}$. In the result the first component is a $s=0$ hole state (linear combinations of the $s=1$ hole states), the second component is a $s=1$ hole state. Combining Eqs. (\ref{A2}) and (\ref{B3}), we write in summary
\begin{eqnarray}
a_\alpha^\dagger a_\beta^\dagger (P^\dagger)^{N-1} |0\rangle  \nonumber \\
= \eta^1_1 \frac{a_{\tilde{\alpha}} a_{\tilde{\beta}}}{v_\alpha v_\beta} (\bar{P})^{\bar{N}-1} |\bar{0}\rangle + \delta_{\beta\tilde{\alpha}} \frac{\eta^1_0}{v_\alpha} (\bar{P})^{\bar{N}} |\bar{0}\rangle .  \label{AB_final}
\end{eqnarray}
Equation (\ref{AB_final}) tells that the $s=1$ particle states can be expressed as the $s=1$ hole states. The converse is also true. Thus the $s=1$ particle space and the $s=1$ hole space are the same.

The $s \ge 2$ states could be treated similarly. In general, an unnormalized $s = h+k$ particle state is written as
\begin{eqnarray}
state = \underbrace{a_{\alpha_1}^\dagger a_{\alpha_2}^\dagger ... a_{\alpha_{2h}}^\dagger}_{2h} \underbrace{P_{\beta_1}^\dagger P_{\beta_2}^\dagger ... P_{\beta_k}^\dagger}_{k} (P^\dagger)^{N-s} |0\rangle ,  \label{state1}
\end{eqnarray}
where $\alpha_1, \alpha_2, ..., \alpha_{2h}$ belong to different pairs of orbits ($P_{\alpha_1}$, $P_{\alpha_2}$, ..., $P_{\alpha_{2h}}$ are all different). The identity (\ref{Talmi}) in the Pauli-blocked subspace $|0^{[\alpha_1...\alpha_{2h},\beta_1...\beta_k]}\rangle$ reads [$(\Omega-2h-k)-(N-s) = \bar{N}-h$]
\begin{widetext}
\begin{eqnarray}
(P^\dagger_{[\alpha_1...\alpha_{2h},\beta_1...\beta_k]})^{N-s}|0^{[\alpha_1...\alpha_{2h},\beta_1...\beta_k]}\rangle  \nonumber \\
= \frac{(N-s)!}{(\bar{N}-h)!} \frac{\prod_{\gamma \in \Lambda} v_\gamma}{v_{\alpha_1}...v_{\alpha_{2h}}v_{\beta_1}...v_{\beta_k}} (\bar{P}_{[\alpha_1...\alpha_{2h},\beta_1...\beta_k]})^{\bar{N}-h}|\bar{0}^{[\alpha_1...\alpha_{2h},\beta_1...\beta_k]}\rangle \nonumber \\
= \frac{\eta^{s}_{h}}{v_{\alpha_1}...v_{\alpha_{2h}}v_{\beta_1}...v_{\beta_k}} (\bar{P}_{[\alpha_1...\alpha_{2h},\beta_1...\beta_k]})^{\bar{N}-h}|\bar{0}^{[\alpha_1...\alpha_{2h},\beta_1...\beta_k]}\rangle . \nonumber
\end{eqnarray}
Therefore Eq. (\ref{state1}) becomes
\begin{eqnarray}
state = \eta^{s}_{h} \frac{a_{\tilde{\alpha}_1} a_{\tilde{\alpha}_2} ... a_{\tilde{\alpha}_{2h}}}{v_{\alpha_1}...v_{\alpha_{2h}}v_{\beta_1}...v_{\beta_k}} ( \bar{P}_{[\beta_1...\beta_k]} )^{\bar{N}-h} |\bar{0}\rangle .  \label{state2}
\end{eqnarray}
Power expanding the right-hand side,
\begin{eqnarray}
( \bar{P}_{[\beta_1...\beta_k]} )^{\bar{N}-h} |\bar{0}\rangle = ( \bar{P} - \frac{P_{\beta_1}}{v_{\beta_1}} - \frac{P_{\beta_2}}{v_{\beta_2}} - ... - \frac{P_{\beta_k}}{v_{\beta_k}} )^{\bar{N}-h} |\bar{0}\rangle  \nonumber \\
= \sum_{0 \le n \le k} \frac{(\bar{N}-h)!}{(\bar{N}-h-n)!} \sum_{ \{\gamma_1...\gamma_n\} \in \{\beta_1...\beta_k\} } \frac{(-)^n P_{\gamma_1}P_{\gamma_2}...P_{\gamma_n}}{v_{\gamma_1}v_{\gamma_2}...v_{\gamma_n}}  \bar{P}^{\bar{N}-h-n}|\bar{0}\rangle ,  \label{s_comps}
\end{eqnarray}
where the summation index $\{\gamma_1...\gamma_n\} \in \{\beta_1...\beta_k\}$ means taking $n$ different elements $\{\gamma_1...\gamma_n\}$ from the set $\{\beta_1...\beta_k\}$, and summing over all possibilities. Consequently Eq. (\ref{state2}) becomes
\begin{eqnarray}
state = \underbrace{a_{\alpha_1}^\dagger a_{\alpha_2}^\dagger ... a_{\alpha_{2h}}^\dagger}_{2h} \underbrace{P_{\beta_1}^\dagger P_{\beta_2}^\dagger ... P_{\beta_k}^\dagger}_{k} (P^\dagger)^{N-s} |0\rangle  \nonumber \\
= \frac{a_{\tilde{\alpha}_1} a_{\tilde{\alpha}_2} ... a_{\tilde{\alpha}_{2h}}}{v_{\alpha_1}...v_{\alpha_{2h}}v_{\beta_1}...v_{\beta_k}} \sum_{0 \le n \le k} \eta^{s}_{h+n}  
\sum_{ \{\gamma_1...\gamma_n\} \in \{\beta_1...\beta_k\} } \frac{(-)^n P_{\gamma_1}P_{\gamma_2}...P_{\gamma_n}}{v_{\gamma_1}v_{\gamma_2}...v_{\gamma_n}}  \bar{P}^{\bar{N}-h-n}|\bar{0}\rangle ,  \label{s_even}
\end{eqnarray}
where we have used $\eta^{s}_{h} (\bar{N}-h)!/(\bar{N}-h-n)! = \eta^{s}_{h+n}$. The result has components of $s' = h + n = h, h+1, ... , s$. Hence the particle states of generalized seniority $2s$ can be expressed as the hole states of $2s$. The converse is also true. This proves the particle-hole symmetry: the $2N$-particle space and the $2\bar{N}$-hole space truncated at arbitrary generalized seniority $S=2s$ are the same [$0 \le s \le \min(N,\bar{N})$]. This symmetry has been tested numerically by the fast algorithm we developed \cite{Jia_2015} and applied \cite{Qi_2016} recently.

Odd-particle systems also have the particle-hole symmetry in generalized seniority: the $(2N+1)$-particle space and the $(2\bar{N}-1)$-hole space truncated at arbitrary generalized seniority $S=2s+1$ are the same [$0 \le s \le \min(N,\bar{N}-1)$]. The actual transformation between the particle and the hole representations is ($s = h + k$)
\begin{eqnarray}
\underbrace{a_{\alpha_1}^\dagger a_{\alpha_2}^\dagger ... a_{\alpha_{2h+1}}^\dagger}_{2h+1} \underbrace{P_{\beta_1}^\dagger P_{\beta_2}^\dagger ... P_{\beta_k}^\dagger}_{k} (P^\dagger)^{N-s} |0\rangle  \nonumber \\
= \frac{- a_{\tilde{\alpha}_1} a_{\tilde{\alpha}_2} ... a_{\tilde{\alpha}_{2h+1}}}{v_{\alpha_1}...v_{\alpha_{2h+1}}v_{\beta_1}...v_{\beta_k}} \sum_{0 \le n \le k} \eta^{s}_{h+n+1}  
\sum_{ \{\gamma_1...\gamma_n\} \in \{\beta_1...\beta_k\} } \frac{(-)^n P_{\gamma_1}P_{\gamma_2}...P_{\gamma_n}}{v_{\gamma_1}v_{\gamma_2}...v_{\gamma_n}}  \bar{P}^{\bar{N}-1-h-n}|\bar{0}\rangle .  \label{s_odd}
\end{eqnarray}

\end{widetext}

\section{J-Scheme Generalized Seniority  \label{sec_J_scheme}}

In the previous section we show that the particle-hole symmetry exists in the generalized-seniority truncated subspaces, and find the transformation between the particle and the hole representations in M-scheme. In this section we assume the rotational symmetry, and write the transformation in (coupled) J-scheme. The single-particle space is generally written as $\{j_1, j_2, ...,
j_D\}$. The pair structure $v_{jm} = v_j$ is independent of the magnetic quantum number $m$. We choose the phase of the time-reversed orbit to be
\begin{eqnarray}
\tilde{a}_{j,m} = (-)^{j-m} a_{j,-m} .
\end{eqnarray}
The tensor $\tilde{a}_{j}$ transforms in the same way as $a^\dagger_j$ under rotation.

In even systems the J-scheme transformation results from coupling the M-scheme transformation (\ref{s_even}) with Clebsch-Gordan coefficients. For components with the maximal generalized seniority,
\begin{eqnarray}
(a_{j_1}^\dagger a_{j_2}^\dagger ... a_{j_{2s}}^\dagger)^{\tau,J} (P^\dagger)^{N-s} |0\rangle  \nonumber \\
= \eta^s_s \frac{(\tilde{a}_{j_1} \tilde{a}_{j_2} ... \tilde{a}_{j_{2s}})^{\tau,J}}{v_{j_1} v_{j_2} ... v_{j_{2s}}} (\bar{P})^{\bar{N}-s} |\bar{0}\rangle + O(s-1) ,  \label{J_tran}
\end{eqnarray}
where $O(s-1)$ represents terms of generalized seniority $2(s-1)$ and less, and $\tau$ collects all the intermediate angular momenta to specify the state in the selected coupling scheme. The result is neat: simply replacing $a^\dagger_{j_i}$ by $\tilde{a}_{j_i}$.

Next we write the full expression for $s=1$ and $2$; namely, give expressions of $O(s-1)$. For $s=1$,
\begin{eqnarray}
(a_{j_1}^\dagger a_{j_2}^\dagger)^J (P^\dagger)^{N-1} |0\rangle = C_{s=1} + \delta_{J0} \delta_{j_1 j_2} \frac{\eta^1_0 \hat{j}_1}{v_{j_1}} (\bar{P})^{\bar{N}} |\bar{0}\rangle ,  \nonumber
\end{eqnarray}
where $\hat{j}_1 \equiv \sqrt{2j_1+1}$, and
we have used $(a_j^\dagger a_j^\dagger)^0_0 = \sum_m a_{jm}^\dagger \tilde{a}_{jm}^\dagger / \sqrt{2j+1}$. $C_{s=1} = \eta^1_1 (\tilde{a}_{j_1} \tilde{a}_{j_2})^J (\bar{P})^{\bar{N}-1} |\bar{0}\rangle / (v_{j_1} v_{j_2})$ stands for the $s=1$ term as given in Eq. (\ref{J_tran}).

For $s=2$, the unpaired part $(a_{j_1}^\dagger a_{j_2}^\dagger a_{j_3}^\dagger a_{j_4}^\dagger)^{\tau,J}$ divides into several cases. If the four particles are on different $j$-levels ($j_1$, $j_2$, $j_3$, $j_4$ are all different), the $O(s-1)$ terms vanish and the full expression is given by Eq. (\ref{J_tran}). If only two of the four $j$'s are the same ($j$, $j_3$, $j_4$ are different),
\begin{eqnarray}
[(a_{j}^\dagger a_{j}^\dagger)^\lambda (a_{j_3}^\dagger a_{j_4}^\dagger)^{\lambda'}]^J (P^\dagger)^{N-2} |0\rangle  \nonumber \\
= C_{s=2} + \delta_{\lambda 0} \delta_{\lambda' J} \frac{\eta^2_1 \hat{j} (\tilde{a}_{j_3} \tilde{a}_{j_4})^{J}}{v_{j} v_{j_3} v_{j_4}} (\bar{P})^{\bar{N}-1} |\bar{0}\rangle ,  \nonumber
\end{eqnarray}
where $\lambda$ is even and $C_{s=2} = \eta^2_2 [(\tilde{a}_{j} \tilde{a}_{j})^\lambda (\tilde{a}_{j_3} \tilde{a}_{j_4})^{\lambda'}]^J (\bar{P})^{\bar{N}-2} |\bar{0}\rangle / (v_{j}^2 v_{j_3} v_{j_4})$ according to Eq. (\ref{J_tran}). If the four $j$'s are pairwise equal ($j \ne j'$),
\begin{eqnarray}
[(a_{j}^\dagger a_{j}^\dagger)^\lambda (a_{j'}^\dagger a_{j'}^\dagger)^{\lambda'}]^J (P^\dagger)^{N-2} |0\rangle  \nonumber \\
= C_{s=2} + \delta_{\lambda 0} \delta_{\lambda' J} (1-\delta_{\lambda' 0}) \frac{\eta^2_1 \hat{j} (\tilde{a}_{j'} \tilde{a}_{j'})^{J}}{v_{j} v_{j'}^2} (\bar{P})^{\bar{N}-1} |\bar{0}\rangle \nonumber \\
+ \delta_{\lambda' 0} \delta_{\lambda J} (1-\delta_{\lambda 0}) \frac{\eta^2_1 \hat{j'} (\tilde{a}_{j} \tilde{a}_{j})^{J}}{v_{j'} v_{j}^2} (\bar{P})^{\bar{N}-1} |\bar{0}\rangle  \nonumber \\
+ \delta_{\lambda' 0} \delta_{\lambda 0} \delta_{J 0} \frac{\eta^2_0 \hat{j} \hat{j'}}{v_{j} v_{j'}} (\bar{P})^{\bar{N}} |\bar{0}\rangle ,  \nonumber
\end{eqnarray}
where $\lambda$, $\lambda'$ are even and $C_{s=2} = \eta^2_2 [(\tilde{a}_{j} \tilde{a}_{j})^\lambda (\tilde{a}_{j'} \tilde{a}_{j'})^{\lambda'}]^J (\bar{P})^{\bar{N}-2} |\bar{0}\rangle / (v_{j}^2 v_{j'}^2)$. If three or four particles are on the same $j$-level, the result is complicated involving various recoupling of the identical $a_j^\dagger$ operators; we skip it here.

Odd-particle systems could be treated similarly. Coupling the M-scheme transformation (\ref{s_odd}) with Clebsch-Gordan coefficients, we have for components of the maximal generalized seniority
\begin{eqnarray}
(a_{j_1}^\dagger a_{j_2}^\dagger ... a_{j_{2s+1}}^\dagger)^{\tau,J} (P^\dagger)^{N-s} |0\rangle  \nonumber \\
= - \eta^s_{s+1} \frac{(\tilde{a}_{j_1} \tilde{a}_{j_2} ... \tilde{a}_{j_{2s+1}})^{\tau,J}}{v_{j_1} v_{j_2} ... v_{j_{2s+1}}} (\bar{P})^{\bar{N}-1-s} |\bar{0}\rangle + O(s-1) .~  \label{J_tran_odd}
\end{eqnarray}

We write the full expression for $S = 2s+1 = 3$; namely, give expressions of $O(s-1)$ for $(a_{j_1}^\dagger a_{j_2}^\dagger a_{j_3}^\dagger)^{\tau,J} (P^\dagger)^{N-1} |0\rangle$. If the three particles are on different $j$-levels ($j_1$, $j_2$, $j_3$ are all different), the $O(s-1)$ terms vanish and the full expression is given by Eq. (\ref{J_tran_odd}). If only two of the three $j$'s are the same ($j \ne j'$),
\begin{eqnarray}
((a_{j}^\dagger a_{j}^\dagger)^\lambda a_{j'}^\dagger)^J (P^\dagger)^{N-1} |0\rangle  \nonumber \\
= C_{s=1} - \delta_{\lambda 0} \delta_{J j'} \frac{\eta^1_1 \hat{j} \tilde{a}_{j'}}{v_{j} v_{j'}} (\bar{P})^{\bar{N}-1} |\bar{0}\rangle ,  \nonumber
\end{eqnarray}
where $C_{s=1} = - \eta^1_2 ((\tilde{a}_j \tilde{a}_j)^\lambda \tilde{a}_{j'})^J (\bar{P})^{\bar{N}-2} |\bar{0}\rangle / (v_j^2 v_{j'})$ according to Eq. (\ref{J_tran_odd}). If the three $j$'s are the same,
\begin{eqnarray}
(a_{j}^\dagger a_{j}^\dagger)^0 a_{jm}^\dagger (P^\dagger)^{N-1} |0\rangle  \nonumber \\
= C_{s=1} - \frac{ (2j-1) \eta^1_1 \tilde{a}_{jm} }{ \hat{j} v_j^2 } (\bar{P})^{\bar{N}-1} |\bar{0}\rangle ,  \nonumber
\end{eqnarray}
where $C_{s=1} = - \eta^1_2 (\tilde{a}_j \tilde{a}_j)^0 \tilde{a}_{jm} (\bar{P})^{\bar{N}-2} |\bar{0}\rangle / v_j^3$. And for even $\lambda \ne 0$ \begin{eqnarray}
(a_{j}^\dagger a_{j}^\dagger)^\lambda_0 a_{jm}^\dagger (P^\dagger)^{N-1} |0\rangle  \nonumber \\
= C_{s=1} + 2 (-)^{j-m} C_{jm j-m}^{\lambda 0} \frac{\eta^1_1 \tilde{a}_{jm}}{v_j^2} (\bar{P})^{\bar{N}-1} |\bar{0}\rangle ,  \nonumber
\end{eqnarray}
where $C_{s=1} = - \eta^1_2 (\tilde{a}_j \tilde{a}_j)^\lambda_0 \tilde{a}_{jm} (\bar{P})^{\bar{N}-2} |\bar{0}\rangle / v_j^3$.

\section{Microscopic Interacting Boson (Fermion) Model  \label{sec_IBM}}

The interacting boson model (IBM) \cite{Arima_1975, Otsuka_1978, Iachello_1987, Casten_1988, IBM_book} uses bosons of various multipolarities as building blocks of the model space. Microscopically, the bosons are identified \cite{Allaart_1988, Otsuka_1978, Iachello_1987, IBM_book} as collective nucleon pairs
\begin{eqnarray}
B^\dagger = \sum_{j_1 j_2} \beta_{j_1 j_2} (a_{j_1}^\dagger a_{j_2}^\dagger)^\lambda  \label{B_def}
\end{eqnarray}
with the multipolarity $\lambda$ and the pair structure $\beta_{j_1 j_2}$. Initially only $\mathcal{S}$ ($\lambda=0$) and $\mathcal{D}$ ($\lambda=2$) bosons are introduced, but later it is found that bosons with $\lambda > 2$ are frequently necessary. [Our $P^\dagger$ operator (\ref{P_dag}) is macroscopically the IBM $\mathcal{S}$ boson]. The mapping from the shell model determines the bosonic Hamiltonian.

It was suspected \cite{Pittel_82} that beyond the half-filled shell the particle-hole ambiguity arises in the mapping. Later the ambiguity was clarified by Talmi \cite{Talmi_82}, Johnson and Vincent \cite{Johnson_85}, in the $\mathcal{S}$-$\mathcal{D}$ model space. Here we show that their conclusion generalizes to larger spaces that consist of many kinds of bosons. From Eq. (\ref{J_tran}) we immediately have
\begin{eqnarray}
(B_1^\dagger B_2^\dagger ... B_s^\dagger)^{\tau,J} (P^\dagger)^{N-s} |0\rangle  \nonumber \\
= \eta^s_s (\bar{B}_1 \bar{B}_2 ... \bar{B}_s)^{\tau,J} (\bar{P})^{\bar{N}-s} |\bar{0}\rangle + O(s-1) ,  \label{micro_IBM}
\end{eqnarray}
where the $s$ particle-pair operators $B_i^\dagger = \sum \beta^i_{j_1 j_2} (a_{j_1}^\dagger a_{j_2}^\dagger)^{\lambda_i}$ ($i = 1,2,...,s$) could have different multipolarities $\lambda_i$ and pair structures $\beta^i_{j_1 j_2}$. The corresponding hole-pair operators are
\begin{eqnarray}
\bar{B}_i = \sum_{j_1 j_2} \frac{\beta^i_{j_1j_2}}{v_{j_1} v_{j_2}} (\tilde{a}_{j_1} \tilde{a}_{j_2})^{\lambda_i} .  \label{Bbar_def}
\end{eqnarray}
In the IBM different types of bosons commute; hence the boson states are microscopically identified as the nucleon-pair states in Eq. (\ref{micro_IBM}) projected onto the maximal generalized seniority $2s$ (for example see Ref. \cite{Allaart_1988}). Therefore the $O(s-1)$ terms drop out; the microscopic IBM model space preserves the particle-hole symmetry. Orthogonalizing for example $(B_1^\dagger B_1^\dagger)^{J} (P^\dagger)^{N-2} |0\rangle$ and $(B_2^\dagger B_2^\dagger)^{J} (P^\dagger)^{N-2} |0\rangle$ (the two different bosons ${\mathcal{B}}_1$ and ${\mathcal{B}}_2$ commute in the IBM) does not affect the conclusion; the orthogonalization happens in the particle and the hole representations simultaneously.

For odd-particle systems, the microscopic interacting boson fermion model (IBFM) \cite{Iachello_1979, IBFM_book} uses the model space consists of one (or more) unpaired fermion and various bosons. From Eq. (\ref{J_tran_odd}) we immediately have
\begin{eqnarray}
(a_j^\dagger B_1^\dagger B_2^\dagger ... B_s^\dagger)^{\tau,J} (P^\dagger)^{N-s} |0\rangle  \nonumber \\
= - \frac{\eta^s_{s+1}}{v_j} (\tilde{a}_j \bar{B}_1 \bar{B}_2 ... \bar{B}_s)^{\tau,J} (\bar{P})^{\bar{N}-1-s} |\bar{0}\rangle + O(s-1) ,  \label{micro_IBFM}
\end{eqnarray}
where $B_i^\dagger$ and $\bar{B}_i$ are still defined by Eqs. (\ref{B_def}) and (\ref{Bbar_def}). The normalization $\eta^s_{s+1}/v_j$ is different when $a_j^\dagger$ is on different $j$-levels. Projecting onto the maximal generalized seniority $S=2s+1$, the $O(s-1)$ terms drop out. The microscopic IBFM model space preserves the particle-hole symmetry.

\section{Nucleon-Pair Approximation  \label{sec_NPA}}

Inspired by the IBM, the nucleon-pair approximation (NPA) \cite{Chen_1997, Zhao_2000, Yoshinaga_2000,Jia_2007,Zhao_2014} further truncates the generalized seniority subspace; the unpaired nucleons are coupled into collective pairs of certain multipolarities (quadrupole, octupole, hexadecapole, ...). The NPA basis is
\begin{eqnarray}
(B_1^\dagger B_2^\dagger ... B_s^\dagger)^{\tau,J} (P^\dagger)^{N-s} |0\rangle  \label{basis_NPA}
\end{eqnarray}
as appeared on the left-hand side of Eq. (\ref{micro_IBM}), where $B_i^\dagger = \sum \beta^i_{j_1 j_2} (a_{j_1}^\dagger a_{j_2}^\dagger)^{\lambda_i}$ is still defined by Eq. (\ref{B_def}). In NPA we diagonalize the exact shell-model Hamiltonian inside the NPA subspace, without mapping onto bosons.

Here we show that in general the particle-hole symmetry is lost in the NPA subspace. As a counterexample, we consider the simplest version of NPA consisting of only $S^\dagger$ [our $P^\dagger$ (\ref{P_dag})] and $D^\dagger$ pairs. From Eq. (\ref{s_even}), the particle state of two $D^\dagger$ coupled to $J=4$ is transformed as
\begin{eqnarray}
(D^\dagger D^\dagger)^{J=4} (P^\dagger)^{N-2} |0\rangle  \nonumber \\
= \eta^2_2 (\bar{D} \bar{D})^{J=4} (\bar{P})^{\bar{N}-2} |\bar{0}\rangle + \eta^2_1 (\bar{G})^{J=4} (\bar{P})^{\bar{N}-1} |\bar{0}\rangle .  \label{DD}
\end{eqnarray}
In the hole representation a new hexadecapole pair $\bar{G} = \sum \beta^G_{j_1 j_2} (\tilde{a}_{j_1} \tilde{a}_{j_2})^{\lambda=4}$ appears, and its structure $\beta^G_{j_1 j_2}$ is completely determined by the structure of $D^\dagger$. The particle-hole symmetry is broken.

Near the half-filled major shell, the NPA should be careful in choosing between the particles and the holes as the degree of freedom; the results are generally different.

\section{Other Truncation Schemes  \label{sec_others}}

In this section we consider the particle-hole symmetry in other popular truncation schemes on top of the generalized seniority. These schemes are frequently used to truncate the shell-model space; here they act in the same way onto the {\emph{unpaired}} nucleons of the generalized seniority subspace (\ref{sen_basis}).

In the multi-$j$ model, we introduce $n_j$ as the number of {\emph{unpaired}} nucleons [particles (holes) in the particle (hole) representation] on the $j$-level. The truncation $n_j \le n^{\max}_j$, where $n^{\max}_j$ are pre-selected integers, preserves the particle-hole symmetry. This is easily proved through Eq. (\ref{s_even}): on the right-hand side the number of unpaired holes $n^{{\rm hole}}_j$ is less than (some $\beta$ index is not selected into the $\gamma$ indices) or equal to (all are selected) the number of unpaired particles $n^{{\rm particle}}_j$ of the left-hand side.

However, following the same argument, the truncation $n_j \ge n^{\min}_j$ ($n^{\min}_j$ are pre-selected integers) breaks the particle-hole symmetry.

Another popular truncation scheme is cutting by mean energies of the basis states. For each basis state $|i\rangle$ in the form (\ref{sen_basis}), we compute $E_i = \langle i | H | i \rangle$, and remove all the states with $E_i > E_{\max}$ ($E_{\max}$ is the energy cutoff). In general this scheme breaks the particle-hole symmetry. As shown in Eq. (\ref{s_even}), some hole states from the right-hand side possibly had higher mean energy than the particle state from the left-hand side.

The particle-hole symmetry in other truncation schemes could be analyzed through the transformations (\ref{s_even}) and (\ref{s_odd}) for even and odd systems.

\section{Conclusions    \label{sec_conclusions}}

In this work we show that the particle-hole symmetry survives the truncation from the full shell-model space to the generalized seniority subspace. The explicit transformations between the states in the particle and the hole representations are provided in both the M-scheme and the J-scheme.

Based on the results, we consider this symmetry in popular theories that could be regarded as further truncations on top of the generalized seniority. Specifically, the microscopic interacting boson (fermion) model preserves the symmetry, while the nucleon-pair approximation breaks it. Other studied truncation schemes are restricting the unpaired nucleon number in each $j$-level, and cutting by the mean energy of the basis states.

Practical calculations frequently truncate the shell-model space due to the dimension limit. Near the half-filled major shell, the results of this work guide the choice between the particle and the hole representations, for truncation schemes related to the generalized seniority. More care is due if the symmetry is broken.

\section{Acknowledgement}

Support is acknowledged from the National Natural
Science Foundation of China No. 11405109, and the Hujiang Foundation
of China (B14004).

\end{document}